\documentclass[sigconf]{acmart}

\def\BibTeX{{\rm B\kern-.05em{\sc i\kern-.025em b}\kern-.08emT\kern-.1667em\lower.7ex\hbox{E}\kern-.125emX}}

\acmSubmissionID{850}

\begin{document}

%
\title[Context-aware Item Representation]{Pre-training of Context-aware Item Representation for \\Next Basket Recommendation}

%
\author{Jingxuan Yang$^1$, Jun Xu$^{2}$, Jianzhuo Tong$^1$, Sheng Gao$^1$, Jun Guo$^1$, Jirong Wen$^2$}

\affiliation{
  \institution{$^1$Beijng University of Posts and Telecommunications, \\$^2$School of Information, Renmin University of China}
}
\email{ {yjx, tongjianzhuo, gaosheng, guojun}@bupt.edu.cn,
 junxu@ruc.edu.cn, jirong.wen@gmail.com}
\begin{abstract}
Next basket recommendation, which aims to predict the next a few items that a user most probably purchases given his historical transactions, plays a vital role in market basket analysis. From the viewpoint of item, an item could be purchased by different users together with different items, for different reasons. Therefore, an ideal recommender system should represent an item considering its transaction contexts. Existing state-of-the-art deep learning methods usually adopt the static item representations, which are invariant among all of the transactions and thus cannot achieve the full potentials of deep learning. Inspired by the pre-trained representations of BERT in natural language processing, we propose to conduct context-aware item representation for next basket recommendation, called Item Encoder Representations from Transformers (IERT). In the offline phase, IERT pre-trains deep item representations conditioning on their transaction contexts. In the online recommendation phase, the pre-trained model is further fine-tuned with an additional output layer. The output contextualized item embeddings are used to capture users' sequential behaviors and general tastes to conduct recommendation. Experimental results on the Ta-Feng data set show that IERT outperforms the state-of-the-art baseline methods, which demonstrated the effectiveness of IERT in next basket representation.  
\end{abstract}

%
%


\keywords{Next basket recommendation, BERT model}

\maketitle

\section{Introduction}
Market basket analysis has been widely used in online shopping companies to help retailers understand the customers' purchase behaviors.
In real-world, a customer usually visits a store multiple times and purchase a set of items as a basket at each of his visit. 
Given his purchase records, how to predict the items a user probably buy in the next visit becomes a crucial task, called next basket recommendation~\cite{lee2005classification-based, wang2014modeling, gatzioura2015a}. 

In fact, an item is purchased in the next visit may because it matches the user's general taste (i.e., what items a user is interested in), or because it matches the user's sequential behaviors 
(i.e., purchasing one item is related to purchasing another). Existing methods for next basket recommendation focus on modeling these two factors. For example, the traditional collaborative filtering (CF)-based methods represent the users' general tastes by factorizing the user-item matrix~\cite{koren2009matrix}. However, the users' sequential transaction behaviors are overlooked. Pattern-based method~\cite{guidotti2017next} models the evolution of customer's purchasing behaviors considering the purchase frequency and the periodic changes. In recent years, deep neural networks have been applied to next basket recommendation. 
Hierarchical Representation Model (HRM)~\cite{wang2014modeling} applies nonlinear operations 
to model the interaction between sequential behavior and users' general taste. Dynamic REcurrent bAsket Model (DREAM)~\cite{yu2016a} improves HRM by adopting RNN to model interactions among apart baskets. Attribute-aware Neural Attentive Model (ANAM)~\cite{bai2018an} further considers item attribute and utilizes attention mechanism to capture user's evolving interests.

Though promising improvements have been observed, existing deep approaches still have limitations. All these neural network methods focus on representing the user's general and transaction specific interests. The items, however, are simply represented with the fixed-length static vectors. 
In real recommendation phenomenon, a user may purchase an item together with different items for different intentions, which correspond to different aspects of the item. For example, a user would like to equip some accessories for his mobile phone when earphone is purchased together with usb-cable. In contrast, the user may also purchase a number of earphones with the intent to do wholesale business. It is obvious that these two intents 
reflect different aspects of earphone. To conduct better recommendation, the earphone representations in these two 
scenarios should also be changed accordingly. Existing deep methods make use of the invariant item representations among all of the transactions (i.e., usually looking up from the transformation matrix), and thus limit the further improvements of 
neural networks.
In this paper, to address this issue and inspired by the success of Bidirectional Encoder Representations from Transformers (BERT) model~\cite{devlin2018bert} in natural language processing (NLP), we propose to represent the items with a fine-tuning based transfer learning architecture. Specifically, in the offline pre-train phase, we first train the model to produce context-aware item representations using sequential transaction corpus.
Then, in the online recommendation phase, we fine-tune the pre-trained parameters and the final contextualized item representations can reflect both the user's local tastes and her global sequential behaviors at the item level.

The proposed model, called Item Encoder Representations from Transformer (IERT), offers several advantages: (1) It produces context-aware item representations which is closer to the nature of next basket recommendation; (2) It employs a two-stage process to capture both the user general tastes from the pre-training and the sequential behaviors from the fine-tuning. Experiments on Ta-Feng showed the superiority of the proposed IERT over the state-of-the-art baselines, indicating the effectiveness of modeling the context-aware item representation in recommendation with pre-traning and fine-tuning.

\section{BERT}
\begin{figure}[t]
  \centering
  \includegraphics[width=\linewidth]{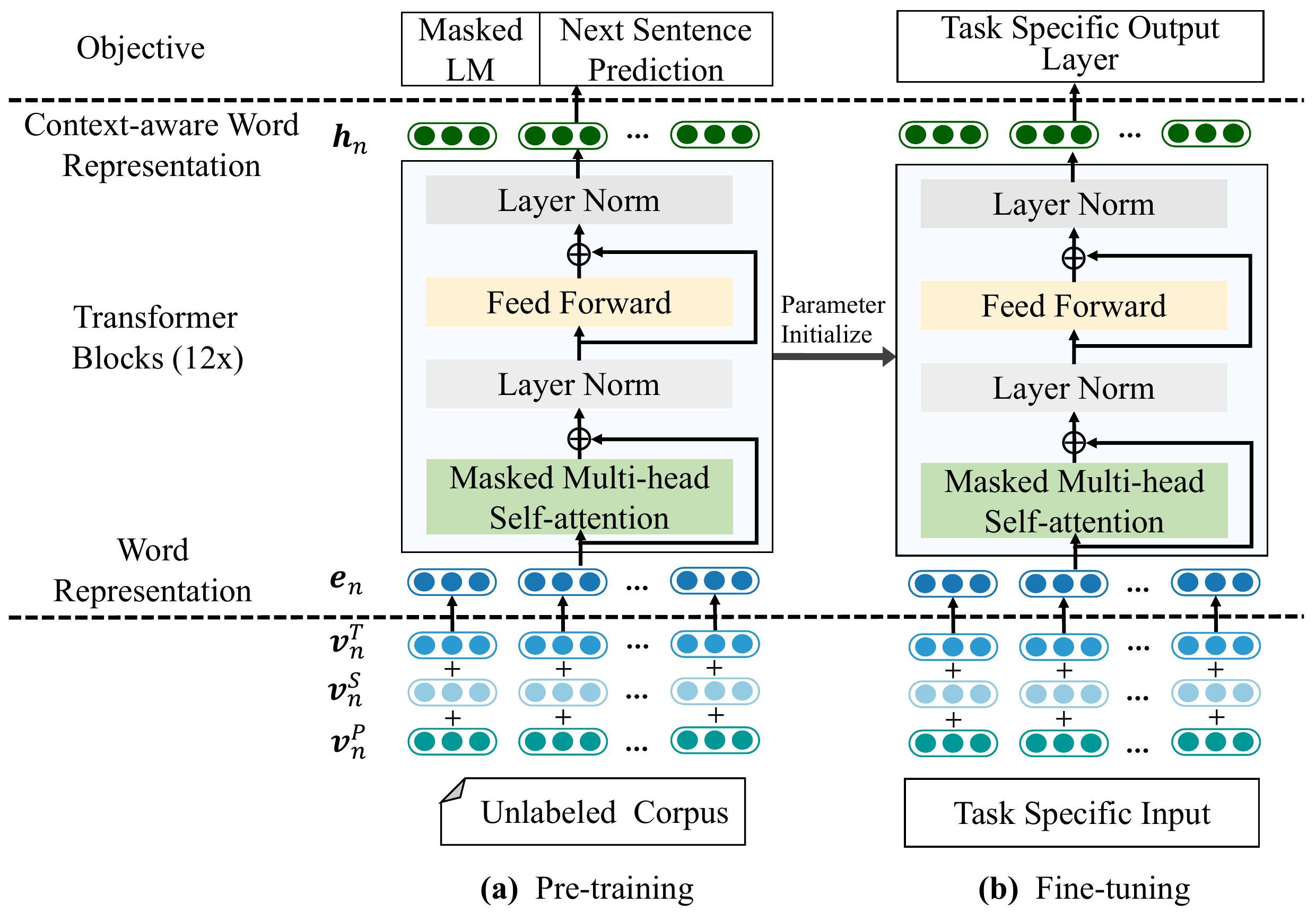}
  \caption{BERT training procedure. (a) shows the pre-training process and Transformer architecture. (b) shows the fine-tuning process, which modifies the pre-trained parameters by supervised target task.}
  \label{bert framework}
\end{figure}

Our proposed model is inspired by the success of BERT in NLP, which aims to encode deep bidirectional language representations.  
In this section, we briefly introduce the 
training procedure of BERT, which is composed of the pre-training stage and the fine-tuning stage as illustrated in Figure~\ref{bert framework}.

\noindent \textbf{Pre-training.} 
BERT tries to learn a deep bidirectional language representation leveraging both left 
and right context. Specifically, given a document-level corpus  $C = \{c_1, c_2, \cdots, c_N\}$ including $N$ tokens, the model first constructs input representation $\textbf{e}_{n}$ for each token $c_n$ by summing the corresponding token embedding $\textbf{v}_n^T$, segment embedding $\textbf{v}_n^S$ (i.e., distinguish two sentences by adding embedding A and B for each token in different sentences), and position embedding $\textbf{v}_n^P$ (i.e., indicate the order of the token in the sentence). Input representations are then fed into a set of Transformer blocks to obtain context-aware token representations. Each Transformer block is composed of a multi-head attention which is followed by a feed-forward layer to obtain an output representation.

The hidden states of the last Transformer block are taken as context-aware representations $\textbf{h}_{n}$ to conduct the pre-training based on two unsupervised prediction tasks: (1) The Masked Language Model (MLM) task, which 
allows the model to predict target masked word 
fusing the left and the right context. 
(2) The next sentence prediction task, which aims to determine the order of two sentences. 


\noindent \textbf{Fine-tuning.} 
BERT adapts the parameters to a supervised target task. Suppose we are given a set of labeled data $L$ as the input, where each instance consists of a sequence of input tokens $x_1,\cdots,x_m$ and a label $y$. The inputs are first passed through the pre-trained model to obtain the last transformer block's activation, 
which is used to make prediction. 
The model 
includes language model likelihood as an auxiliary objective to improve generalization of the supervised model by alleviating the bias on target task and accelerate convergence 
~\cite{rei2017semi-supervised}.

BERT has shown its effectiveness in a variety of NLP tasks including general language understanding, question answering, named entity recognition and grounded commonsense inference. In this paper, we propose to adapt the BERT model for the task of next basket recommendation. 

\section{\mbox{Context-aware Item Representation}}
We propose to adjust the representation mechanism in BERT to produce context-aware item representation and apply the modified one to next basket recommendation.

\subsection{Problem Analysis}
Suppose that we have a set of users $U = \{u_1, u_2, ..., u_{|U|}\}$ and items $I = \{i_1, i_2, ..., i_{|I|}\}$, where the total number of users and items are $|U|$ and $|I|$, respectively. Given a user $u$, his/her historical transactions $B^u$ are composed of a sequence of baskets $\{B_1^u, B_2^u, \cdots, B_t^u\}$ sorted in chronological order, where $t$ denotes the time step and $B_t^u \subseteq I$. The purchase history of all users is denoted as $B = \{B^u\}_{u=1}^{|U|}$.
The goal of next basket recommendation is to predict the items that the user $u$ would probably purchase in his next visit, given his 
historical records. The problem can be reformulated as making a personalized ranking among all items for each user at time step $t+1$. The top $K$ items are recommended to the user from the ranking list. 

Recently, deep neural networks have been employed to solve the next basket recommendation task. 
These methods explore users' general taste 
by modeling user-item interactions and transaction specific interests 
by modeling item-item interactions. Usually, a lookup layer is utilized to represent each item $i$ in the item set as a static vector ${\mathbf{v}}_i$ with fixed-length as:
\begin{equation}
{\mathbf{v}}_i = {\rm LOOKUP}({\mathbf{P}}, i),
\end{equation}
where ${\mathbf{P}} \in \mathbb{R}^{D \times |I|}$ denotes the transformation matrix of items, and $D$ is the length of the vector.

Though promising results have been achieved, existing approaches still have limitations. In most cases, a user purchases an item along with different items due to different intentions. For example, a user would like to obtain bulk purchase discounts when the recipe book is bought together with math book and story book, while she attempts to cook a dish when the recipe book is bought together with tomato and olive oil. Obviously, the recipe book is purchased 
utilizing its different functions, and should have different representations accordingly. Neural network is good at representing the semantics 
by latent vectors and modeling interactions as an universal approximator. Thus accurate representations are important to improve the performances of neural network methods. However, existing 
methods express the same item with an unvarying vector ${\mathbf{v}}_i$ among all transactions, and cannot achieve the full potentials of deep neural networks. 

\subsection{Our Approach: IERT}
\begin{figure}[t] 
  \centering
  \includegraphics[width=\linewidth]{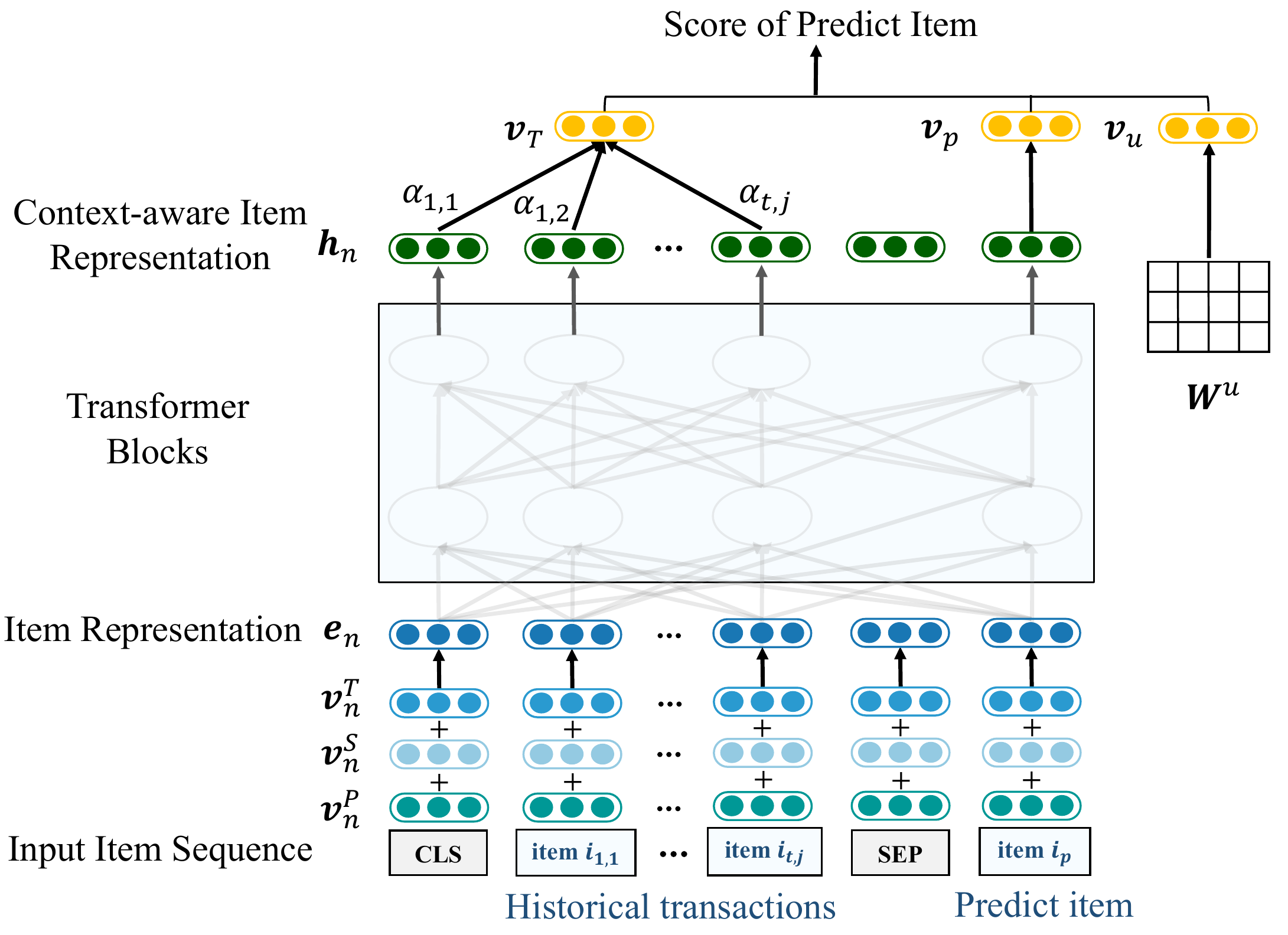}
  \caption{Architecture of IERT fine-tuning process, which conducts recommendation using context-aware item representations $\mathbf{h}_n$. `CLS' and `SEP' are the special symbols for indicating the beginning of the sequence and separating the historical transactions and the predict item, respectively.}\label{fig:IERT}
\end{figure}

To address the above issues, we propose to adapt the BERT to produce context-aware item representations, called IERT, 
which is shown in Figure~\ref{fig:IERT}. Specifically, IERT regards each item as a word in BERT, each basket as a sentence, and a sequence of baskets of the same user as a document. Similar to that of 
BERT, the learning procedure of IERT also consists of the pre-training stage and fine-tuning stage. 

\subsubsection{Pre-training Stage}
Given the 
sequential transaction records $B$ for a set of users, we use $B_t^u$ denote the items $\{i_{t,1}^u, i_{t,2}^u, \cdots, i_{t,j}^u\}$ user $u$ purchased in the $t$-th visit. 
The input representation of each item is constructed by summing three kinds of embeddings, similar to that of input representation in BERT. To adapt for next basket recommendation, the original two pre-training tasks in BERT are modified. Specifically, 
the MLM task is adjusted as Masked Item Prediction task to predict the vocabulary ID of randomly masked item $i_{t,k}^u$ based on the rest of the items 
in the same basket. The modification allows the model to produce context-aware item representations as well as explore users' local transaction behaviors. The objective function of this task is the likelihood of the transaction records:
\begin{equation}
\resizebox{0.923\hsize}{!}{$
L_1(B) = \sum_{u}\sum_{t}\sum_{k}\log P(i_{t,k}^u|i_{t,1}^u,...,i_{t,(k-1)}^u,i_{t,(k+1)}^u,...,i_{t,j}^u;\Theta),
$}
\end{equation}
\indent The next sentence prediction task is also adjusted as Next Basket Prediction to make our model understand the order of two baskets as well as explore sequential features among apart baskets. 
The objective function becomes:
\begin{equation}
L_2(B) = \sum_{i \in I}\sum_{t}\delta(i_{t+1}|i_t) \log(P(i_{t+1}|i_t)),
\end{equation}
where $\delta(i_{t+1}|i_t)$ denotes the annotated label of whether $i_{t+1}$ is the next basket of $i_t$. 
In this work, we construct the basket pairs consecutively or apart with $50\%$ chance respectively.

The overall pre-training loss is the sum of the 
masked item prediction likelihood and the 
next basket prediction likelihood:
\begin{equation}\label{eq:pretrain}
L_3(B) = L_1(B) + L_2(B).
\end{equation}

\subsubsection{Fine-tuning Stage}
After pre-training the model with objective in Eq.~(\ref{eq:pretrain}), the model parameters are fine-tuned at each recommendation. That is, the output states of the last Transformer block are used as the context-aware item representations to explore both users' sequential behaviors and general tastes. 

Formally, given a user $u$, an input instance of the fine-tune stage is a sequence of historical transactions $B^u$ and a candidate item $i$ which probably be purchased in the next visit. Instead of constructing a representation for each basket, IERT models the purchase records at the fine-grained item level. Therefore, the historical transactions of user $u$ can be further presented as sequentially combining the items in each basket $B^u = \{i_{1,1}^u, i_{1,2}^u, ..., i_{t,j}^u\}$.

The history transactions $B^u$ and the predict item $i$ can be packed together as a single sequence, separating with a special token ([SEP]). Then the sequence is fed into the same transformer model as in the pre-training stage. As a result, the output hidden states
\begin{math}
  {\mathbf{H}} = \{{\mathbf{h}}_{1,1}, {\mathbf{h}}_{1,2}, \cdots, {\mathbf{h}}_{t,j} \}
\end{math}
and ${\mathbf{h}}_i$ are token as context-aware representations for historical transaction items and the predict item.

Attention mechanism is employed to capture the global and the local sequential behaviors from fine-grained item level, through constructing the representation of historical transactions. For the predict item ${\mathbf{h}}_i$, the historical transaction is presented as: 
\begin{equation}
{\mathbf{v}}_{B} = \sum_{t=1}^{t} \sum_{j=1}^{|{\mathbf{h}}_t|} \alpha_{t,j} \cdot {\mathbf{h}}_{t,j},
\end{equation}
where $\alpha_{t,j}$ is defined as:
\begin{equation}
\alpha_{t,j} = \frac{{\rm exp}({\mathbf{W}}^{1 \times D}({\mathbf{h}}_i \odot {\rm\textbf{h}}_{t,j}) + b^1)}{\sum_{{t'}=1}^{t} \sum_{{j'}=1}^{|{\mathbf{h}}_t|} {\rm exp}({\mathbf{W}}^{1 \times D}({\mathbf{h}}_i \odot {\mathbf{h}}_{t',{j'}}) + b^1)}
\end{equation}
\indent The index of user $u$ is transformed to an latent vector through a lookup layer:
\begin{equation}
{\mathbf{v}}_u = {\text{LOOKUP}}({\mathbf{Q}}^T, u),
\end{equation}
where ${\mathbf{Q}} \in \mathbb{R}^{D \times |U|}$ denotes the transformation matrix for lookup.

\subsubsection{Online recommendation}
Given a user $u$ and his historical transactions $B^u$, the probability of an item $i$ being purchased in the next visit is calculated by softmax function:
\begin{equation}
    p(i \in B_{t+1}^u|u, B_{1, t}^u) = \frac{{\rm exp}({\mathbf{h}}_i^T \cdot ({\mathbf{v}}_u \odot {\mathbf{v}}_{B}))}{\sum_{{i'}=1}^{|I|} {\rm exp}({\mathbf{h}}_{i'}^T \cdot ({\mathbf{v}}_u \odot {\mathbf{v}}_B))},
\end{equation}
where ${\mathbf{v}}_u \in \mathbb{R}^{D \times 1}$ is the vector representation of user $u$, ${\mathbf{v}}_B \in \mathbb{R}^{D \times 1}$ is the context-aware 
transaction representation.

In the learning process of IERT, weighted cross-entropy is employed as the objective function:
\begin{equation}
    L=\sum_{u \in U} \sum_{B_t^u \in B^u} \sum_{i \in i_t} (-m \cdot y_i \cdot \log p_i - n \cdot (1-y_i) \cdot \log (1-p_i)),
\end{equation}
where $p_i$ is the probability of an item $i$ purchased in the next visit and $y_i$ denotes the annotated label of item $i$, that is, $y_i=1$ if it is purchased in the next visit, otherwise 0.
\subsection{Differences from BERT}
IERT is inspired by the BERT model in NLP. In that sense, it is similar to BERT and share a number of merits from BERT. However, it also has several striking differences from BERT:

First, in the pre-training stage, to learn context-aware representations and explore sentence relationships, BERT takes the order information of the words and the sentences into consideration. In next basket recommendation, however, the order among the transactions is important while the items in the same transaction were bought without strict order. 
Based on the observation, IERT modifies the pre-training objective so as to make the training focus on modeling the order information among transactions.

Second, in the fune-tuning stage, BERT usually receives a pair of sentences in order to explore relationships between them. IERT, however, aims to build intention-related transaction representations, and can only receive historical transactions before current time step. 

Third, BERT usually leverages various large-scale datasets as the pre-training corpus because the same word in different datasets still holds the similar meaning. In next basket recommendation, however, the same item ID in different datasets could represent totally different items and the items are rarely overlapped. Thus, the context-aware item representation is a more challenge task than the word representation task in NLP. 

\vspace{-0.1in}
\section{Experiments}
\textbf{Datasets.} We tested the performances of IERT on Ta-Feng\footnote{\url{http://www.bigdatalab.ac.cn/benchmark/bm/dd?data=Ta-Feng}} data. In Ta-Feng, each basket consists of the items purchased together by one user at a visit. The data set contains $464,118$ transactions belonging to $9,238$ users and $7,793$ items. All the items purchased by less than $10$ users and users purchased less than $10$ items in total were removed to eliminate the noise. In the experiments, the dataset was split into three non-overlapping sets. The last basket of each user is taken as testing set, the penultimate basket is reserved as a held-out validation set for tuning the parameters, and all the remaining baskets are taken as training set.

\noindent \textbf{Experimental Settings.}
Following the practices in~\cite{devlin2018bert}, the proposed IERT model was implemented as follows: the training was conducted with the batch size of $32$ sequences for $40,000$ steps where the original sequences were truncated such that the max number of items in the same basket is $100$. Adam with learning rate of $0.00002$ 
was utilized to conduct the optimization. 
As for the model size, ${\rm BERT_{BASE}}$ structure according to~\cite{devlin2018bert} was chosen, 
where hidden size $H$, the number of Transformer blocks $L$ and the self-attention heads $A$ were set to $768$, $12$,  and $12$, respectively.

Several state-of-the-art next basket recommendation methods were chosen as the baselines, including conventional methods of TOP, NMF~\cite{lee2000algorithms}, and FPMC~\cite{rendle2010factorizing}, and deep methods of HRM~\cite{wang2014modeling}, DREAM~\cite{yu2016a}, and ANAM~\cite{bai2018an}. To test the effectiveness of pre-training mechanism in the context-aware item representations, we compare our IERT with its simplified version which the pre-training stage was removed, denoted as ``IERT (w/ pre-training)''.

\noindent \textbf{Evaluation metrics.}
Same as~\cite{wang2014modeling, yu2016a, bai2018an}, the top K items (K=5) from the ranking list of all items 
were recommended 
to each user $u$. The performances were evaluated with the F1-score and Normalized Discounted Cumulative Gain (NDCG). 


\begin{table}
  \caption{Performance comparison of different methods.}
  \vspace{-0.05in}
  \label{result}
  \begin{tabular}{c|cc}
    \hline
    Model & F1-score@5 & NDCG@5 \\ \hline \hline
    TOP & 0.051 & 0.084 \\ \hline
    NMF & 0.052 & 0.072 \\ \hline
    FPMC & 0.059 & 0.087 \\ \hline
    HRM & 0.062 & 0.089 \\ \hline
    DREAM & 0.133 & 0.173 \\ \hline
    ANAM & 0.146 & 0.190 \\ \hline \hline
    IERT (w/ pre-training) & 0.150 & 0.194 \\ \hline
    IERT & \textbf{0.213} & \textbf{0.340} \\
  \hline
\end{tabular}
\vspace{-0.15in}
\end{table}

\noindent \textbf{Results and analysis.} Results are presented in Table~\ref{result} and boldface indicates the highest number among all of the methods. We can see that the simplified version of our model, i.e., IERT (w/ pre-training), outperformed all of the baseline methods, showing the effectiveness of item-level interaction modeling by transformer encoder. The baseline methods utilize all items in the same transaction to build basket representation. It leads to semantic confusion when some individual items are purchased have nothing to do with others. For example, a user could put toothpaste in the same basket with beer and bread since it is sold at a discount. 

The results in Table~\ref{result} also show IERT worked better than IERT (w/ pre-training), indicating the importance of the pre-training stage in IERT. Compared with the bast baseline ANAM, IERT gained the improvements of $45.9\%$ and $78.9\%$ in terms of F1-score@5 and NDCG@5, respectively, indicating the effectiveness of context-aware item representations in next-basket recommendation. 

\vspace{-0.1in}
\section{Conclusions}
In this paper, we propose to adapt the BERT model in NLP to improve the performances of next basket recommendation, through producing context-aware item representations. The model, referred to as IERT model, first pre-trains the model parameters on the historical purchase transactions and then fine-tunes the model during the online recommendation. Experimental results on publicly available dataset show that IERT outperformed the state-of-the-art baselines, indicating the effectiveness of context-aware item representations. 
\vspace{-0.1in}

%
\bibliographystyle{ACM-Reference-Format}
\bibliography{sample-base}

\end{document}